# Two-Dimensional β-PdX$_2$ (X = S, Te) Monolayers for Efficient Solar Energy Conversion Applications

Mukesh Jakhar and Ashok Kumar[*]

*Department of Physics, School of Basic Sciences, Central University of Punjab, Bathinda, 151401, India*

(January 29, 2022)

*Corresponding Author:

ashokphy@cup.edu.in (Ashok Kumar)

**Abstract**

The search for highly effective and environmentally safe photocatalysts for water splitting and photovoltaic solar cells is essential for renewable solar energy conversion and storage. Based on first principles calculations, we show that novel 2D β-PdX$_2$ (X = S, Te) monolayer possesses excellent stabilities and great potentials in solar energy conversion applications. Comprehensive studies show that the β-PdS$_2$ monolayer exhibits semiconductor characteristics with an indirect gap, suitable band alignment, efficient carrier separation, and high solar to hydrogen (STH) efficiencies, supporting its good photoelectronic performance. The surface catalytic and adsorption/intercalation energies calculation reveals that the photogenerated holes have adequate driving forces to render hydrogen reduction half-reactions to proceed spontaneously and the ability to cover and incorporate water molecules on β-PdS$_2$ monolayer. Besides, the β-PdTe$_2$ monolayer is promising donor material for excitonic solar cells with high photovoltaic performance. More importantly, due to suitable donor band gap and small conduction band offset in the proposed type-II heterostructure, the calculated power conversion efficiencies (PCE) is calculated up to ~23% (β-PdTe$_2$/WTe$_2$), ~21% (β-PdTe$_2$/ MoTe$_2$) and ~18% (β-PdTe$_2$/β-PdS$_2$), making it a promising candidate for solar energy conversion applications.

# 1. Introduction

Solar energy conversion technologies, such as photocatalytic water splitting and photovoltaic cells are demanding to produce clean energy and to reduce significant environmental and energy concerns. Since the realization of semiconductor photocatalyst $TiO_2$ for solar water splitting[1], researchers have devoted significant efforts to explore effective photocatalysts capable of producing hydrogen energy.[2-7] To achieve cost-effective hydrogen production, photocatalysts' solar to hydrogen (STH) efficiency should be at least 10% for large-scale production.[8] The favorable hydrogen and oxygen evolution reaction (HER and OER), efficient separation with fast carrier migration and robust light-harvesting, leads to high STH efficiency. Also, the band gap must be equal or greater than 1.23 eV for a viable photocatalyst, and the VBM and CBM appropriately straddling the redox potentials of the water. If these criteria are satisfied, water can be split without the need of any sacrificial agents.[9] However, due to low quantum efficiency induced by charge recombination both on the surface and in the bulk of the catalysts, only a few shows good photocatalytic activity for water splitting.[10]

In the last few decades, 2D materials have been suggested as promising solar energy harvesters due to high optical absorption, high carrier mobility and sufficient reactive sites[11-19]. Unfortunately, many of them have poor long-term endurance and are effectively inactive for photocatalysis water splitting.[9, 20] Also, the significant energy loss produced by the high overpotential in HER and OER need to be reduced. Therefore, novel 2D photocatalysts, particularly those with low overpotential, are of great interest for realizing the water splitting reaction without the use of sacrificial reagents or co-catalysts.

Among 2D materials, transition-metal dichalcogenides (TMDs) research has sparked a lot of interest in the creation of few-layer or monolayer nanosheets.[21, 22] These advancements bring new life to photocatalytic water splitting due to the fact that TMDs possess a high optical absorption efficiency, high carrier mobility, high charge transfer ability and large catalytic sites.[23-25] Despite the potential advantages of 2D chalcogenides for photocatalysis, the hydrogen yield from water splitting remains low due to the rapid recombination of photogenerated holes and electrons.[26] TMDs with abundant active sites can enhance the efficiency to trigger the overall water splitting reactions.

Besides photocatalyst, the solar cell is also an essential renewable energy conversion technology that converts solar radiation directly into electricity. Solar cell materials with enhanced power conversion efficiency (PCE) have been investigated continually since the invention of the first operational solar cell in 1954.[27] For van der Waals (vdW) heterostructure solar cells with a high PCE should meet two main criteria.[28] First, acceptor and donor materials should have type II band alignments in the heterostructure. Second, the donor material's band gap should be between 0.90-1.70 eV to ensure effective light-harvesting performance. The light-harvesting performance, the carrier mobility and the separation efficiency of excitons also play essential roles in determining the PCE.[29]

Thus, 2D materials also have been regarded as promising candidates for excitonic solar cells (XSCs) due to their adjustable bandgap and good light-harvesting performance with high carrier mobility.[30-32] The electronic characteristics of individual monolayers can be tuned by linking their band edges using a van der Waals (vdW) heterostructure generated by stacking distinct monolayer materials.[33, 34] Therefore, the fabrication of efficient type-II vdW heterostructures is considered promising for next-generation solar cell applications.

In this work, density functional theory (DFT) computations were performed to describe the photocatalytic water splitting and photovoltaic activity of the β-$PdS_2$ and β-$PdTe_2$ monolayers, respectively. We perform molecular dynamics calculations to investigate these monolayers' thermodynamic and kinetic stabilities. Remarkably, the β-$PdS_2$ monolayer has an indirect band gap (2.10 eV) with suitable band edge positions, and the potential to meet the requirement for redox potentials of photocatalysis. The adsorption/intercalation energies are obtained to check whether the coverage and incorporation of water molecules are energetically feasible or not. Furthermore, the β-$PdTe_2$ monolayer exhibits an indirect band gap of ~1.29 eV with suitable type-II vdW heterostructure alignments, suggesting that it could serve as suitable donor material for constructing 2D heterostructure as an excitonic solar cell. Significantly, the estimated PCE of β-$PdTe_2$/β-$PdS_2$, β-$PdTe_2$/$MoTe_2$ and β-$PdTe_2$/$WTe_2$ heterojunctions, can be as high as 17.79%, 20.59% and 23.14% respectively.

**2. Computational method**

The calculated results were obtained by employing density functional theory (DFT) as the Quantum-ESPRESSO package.[35, 36] We employed the the Perdew−Burke−Ernzerhof (PBE) approximation of the exchange-correlation (XC) functional. The convergence criterion for the forces was set to 0.01 eV/Å and a norm-conserving pseudopotential with a 90 Ry energy cutoff. The first Brillouin zone (BZ) integration was performed with 24 × 24 × 1 k-point grid for Monkhorst pack mesh. To break unphysical interactions between periodic system images, we inserted a large vacuum of 20 Å perpendicular to 2D structure. Moreover, the sophisticated HSE06 hybrid functional was also used for calculating the electronic structures[37, 38]. To include van der Waals interactions, we use Grimme's DFT-D2 approach for all heterostructure calculations.[39] The AIMD simulations were performed with Nosé thermostat algorithm at a different temperature of 300 K, 500K and 1000K for a total of 5 ps with a time step of 3 fs using the SIESTA package.[40]

We used the $G_0W_0$+BSE method as incorporated in the YAMBO code[41] coupled with the Quantum-ESPRESSO package for optical absorbance. To achieve convergence, 200 bands are undertaken in the sum over the state for the correlation part of self-energy. To ensure the absorption spectrum converges, three valence and two conduction bands are used to solve the Bethe-Salpeter equation and dielectric function. Moreover, to account for solvent effects in simulations, we used the implicit solvent model within the Environ code[42] of Quantum ESPRESSO using the 78.3 dielectric constant for liquid water. The details of free energy and thermodynamic oxidation and reduction potentials calculations are summarized in the ESI.

## 3 Results and Discussion

### 3.1 Geometric Structure and Stability

The bulk β-PdX$_2$ (X = S, Te) exhibits trigonal symmetry with P3$_1$21 space group[43]. As shown in Fig. 1(a), the layers of β-PdX$_2$ are stacked together along the z-axis through interlayer vdW bonding. The atoms of each layer of bulk β-PdX$_2$ are located in four, and six-membered ring structures parallel to the x-axis as helical chain, where each Pd atom is covalently bound to the four neighboring X (X = S, Te) atoms within the chain, and the chains pile up in the y-direction. Moreover, isolating an individual layer from the β-PdX$_2$ bulk yields the structure of the β-PdX$_2$ monolayer, as shown in Fig. 1(b). The optimized lattice parameters for both β-PdS$_2$, β-PdTe$_2$ monolayer are slightly smaller than the bulk phase. Detailed optimized geometric properties for β-PdX$_2$ bulk and monolayer are summarized in Table S1.

Experiments have shown that mechanical exfoliation can yield various 2D materials with weak interlayer interactions from their bulk counterparts.[44, 45] Here, to explore the possibility of fabricating the energetically favorable β-PdX$_2$ (X = S, Te) from the surface of its layered bulk crystal, we compute their cleavage energy (E$_{cl}$) from a 5-layer slab model with the function of separation distance. Note that this method is highly accurate to calculate cleavage energy comparable to experiments. For example, calculated cleavage energy of graphene (0.33 J/m$^2$)[46] is in excellent agreement with the measured value of cleavage energy (0.32 ± 0.03 J/m$^2$).[47] Remarkably, the theoretical demonstration of the possibility of mechanical exfoliation of Ca$_2$N monolayer with cleavage energy 1.09 J/m$^2$ [48] was recently confirmed in experiments with cleavage energy 1.11 J/m$^2$.[49] Recently, many 2D monolayers such as PdSeO$_3$,[12] BiFeO$_3$,[50] SnP$_3$,[51] and GeTe[52] have been theoretically demonstrated to be potential candidates for experimental fabrications and have also been experimentally synthesized.[53-56]

As depicted in Fig. 1(c), the E$_{cl}$ are 0.47 and 0.88 J/m$^2$ for β-PdS$_2$ and β-PdTe$_2$ monolayers, respectively. Remarkably, the calculated E$_{cl}$ of β-PdS$_2$ is quite comparable with 2D MoS$_2$ (0.42 J/m$^2$)[57], PdSeO$_3$ (0.42 J/m$^2$)[12] crystal that has been realized experimentally via exfoliation techniques. Also, the E$_{cl}$ of β-PdTe$_2$, is lower than that of α-BS (0.96 J/m$^2$)[58], Ca$_2$N (1.08 J/m$^2$)[48], GeP$_3$ (1.14 J/m$^2$)[59] and InP$_3$ (1.32 J/m$^2$)[60]. The cleavage energy calculations suggest the feasibility of experimental fabrication of these monolayers.

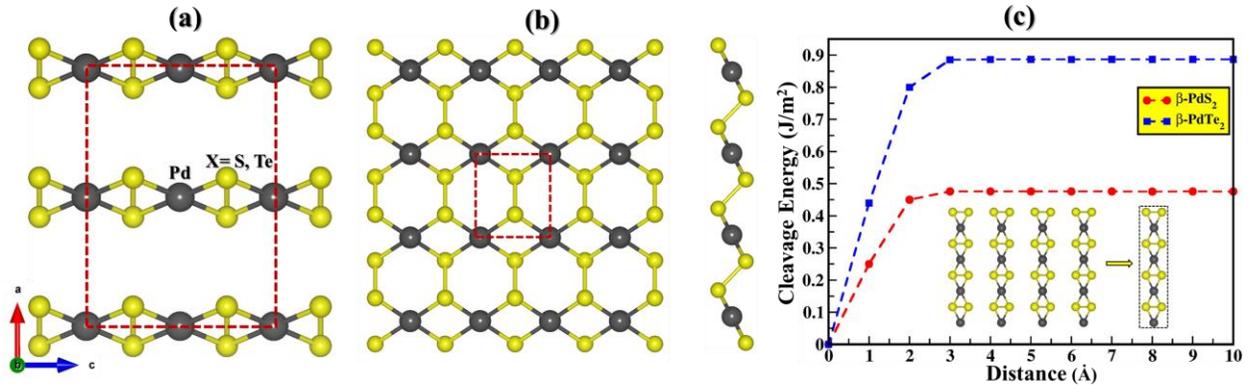

**Fig 1.** (a) The atomic structure of β-PdX$_2$ (X= S, Te) bulk and (b) the top and side views of β-PdX$_2$ (X= S, Te) monolayer. The red dotted line indicates the unit cell. (c) Cleavage energy as a function of the separation distance for β-PdS$_2$ and β-PdTe$_2$ monolayer and four-layer slab (inset). Black: Pd atom. Yellow: Se atom.

Furthermore, to examine the energetic stability of β-PdX$_2$ monolayer, we calculate the cohesive energy (E$_{Coh}$), which is defined as $E_{Coh} = (E_{Pd} + E_X − E_{PdX_2})/3$, where $E_{Pd}$, $E_X$ and $E_{PdX_2}$ are the total energy of a single Pd, X (X=S, Te) atom and total energy of PdX$_2$ unit cell, respectively. The obtained cohesive energy of 4.0 eV/atom for β-PdS$_2$ is higher than that of other 2D materials, such as silicene (3.98 eV/atom), whereas the calculated $E_{Coh}$ of β-PdTe$_2$ (3.3 eV/atom) is comparable to that of black phosphorene (BP) (3.4 eV/atom)[61]. Our investigation demonstrates the high feasibility of exfoliating β-PdX$_2$ (X =S, Te) monolayers from their bulk counterpart as free-standing monolayers.

Next, the dynamic stability of the β-PdX$_2$ monolayers are assessed by examining the phonon spectra along with the high-symmetry points in the Brillouin zone, as shown in Fig. S1(a), S2(b) ESI. No significant imaginary phonon modes are found in the spectrum, revealing the kinetic stability for both β-PdS$_2$ and β-PdTe$_2$ monolayers. Moreover, the thermal stability of β-PdX$_2$ monolayer is also examined by performing AIMD simulations with a relatively larger 4 × 4 supercell at 300 K with a time step of 3 fs. The small total energy fluctuations with time suggest that both β-PdS$_2$ and β-PdTe$_2$ monolayers are thermally stable up to 5000 fs in simulation (Fig. S1(b), S2(b) ESI). The AIMD simulations are also performed at higher temperatures of 500 and 1000 K (Fig. S1(c-d) and S2(c-d) ESI), which reveals good thermodynamically stability of β-PdX$_2$ monolayers, thereby, show promises to future synthesis of these monolayers.[62]

### 3.2 Optoelectronic Properties

After investigating the intrinsic stability and the experimental feasibility, we then computed the optoelectronic properties for the β-PdX$_2$ (X= S, Te) monolayer. As depicted in Fig. 2(a-b), β-PdS$_2$ and β-PdTe$_2$ monolayers are indirect band gap semiconductors of 2.10 eV (1.14 eV) and 1.29 eV (0.65 eV) at the HSE06 (GGA-PBE) level of theories. The VBM is located at near Y (0, 0.5, 0) point for both β-PdS$_2$ and β-PdTe$_2$ monolayers, whereas the CBM lies between the M to Y and Γ to X point in the brillouin zone for β-PdS$_2$ and β-PdTe$_2$ monolayers, respectively. The density of states (DOS) analysis reveals that the states near the Fermi levels are contributed by the mixed of p-orbitals of S/Te and d-orbitals of Pd (Fig. 2(a-b)). The spin-orbit coupling (SOC) effect on the electronic band structure is examined at the PBE+SOC level of theory. Our results reveal that the effect of SOC on β-PdS$_2$ (Fig S3 ESI) is negligible, whereas slight band gap reduction (ΔE$_g$=0.25 eV) has been obtained for β-PdTe$_2$ (Fig S4 ESI).

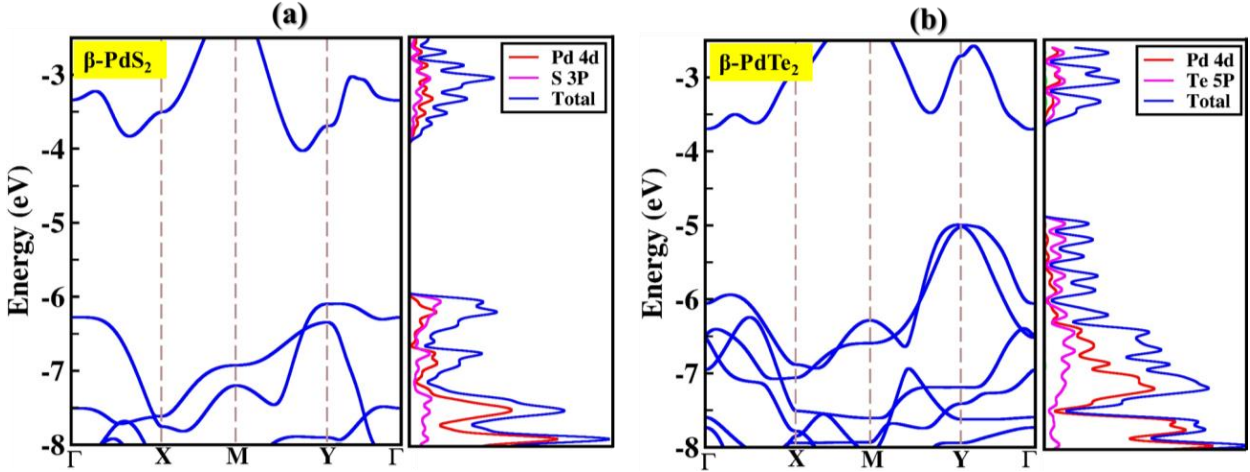

**Fig. 2** The band structure (left) and corresponding density of states (right) of (a) β-PdS$_2$ (b) β-PdTe$_2$ monolayers calculated at HSE06 functional level of theory with vacuum level correction.

### 3.2.1 Carrier Mobilities

In order to explore the separation and migration ability of photoexcited carriers, we calculate the carrier mobilities for β-PdS$_2$ and β-PdTe$_2$ using deformation potential (DP) theory according to phonon-limited scattering model[63] (see ESI). According to our calculated results, the β-PdS$_2$ monolayer has isotropic carrier mobility for electron ($1.04 \times 10^3 \text{cm}^2\text{V}^{-1}\text{S}^{-1}$ and $1.06 \times 10^3 \text{cm}^2\text{V}^{-1}\text{S}^{-1}$) and hole ($2.5 \times 10^1 \text{cm}^2\text{V}^{-1}\text{S}^{-1}$ and $2.7 \times 10^1 \text{cm}^2\text{V}^{-1}\text{S}^{-1}$) along x- and y-direction (Table S2), while β-PdTe$_2$ monolayer has anisotropic values for the electron ($1.25 \times 10^3 \text{cm}^2\text{V}^{-1}\text{S}^{-1}$ and $1.6 \times 10^1 \text{cm}^2\text{V}^{-1}\text{S}^{-1}$) and hole ($1.0 \times 10^1 \text{cm}^2\text{V}^{-1}\text{S}^{-1}$ and $5.4 \times 10^1 \text{cm}^2\text{V}^{-1}\text{S}^{-1}$) along x and y-direction (Table S2). This difference in the nature of carrier mobility of two monolayers can be quantify in terms of the difference in electronegative and orbital radii between S and Te atoms that results into different dispersion of valance and conduction bands near the Fermi level for these monolayers. Different dispersion of bands further leads to change in effective masses and deformation potential along x- and y-direction. For example, effective mass (0.27 and 1.56) and deformation potential ($E_i = 2.0$ eV and 3.38 eV) of electron along x- and y-direction leads to highly anisotropic value of electron mobility in β-PdTe$_2$ monolayer. The electron mobility ($\mu_e$) of β-PdX$_2$ (X=S, Te) monolayers are moderately large and much higher than that of MoS$_2$ ($72.16 \text{ cm}^2\text{V}^{-1}\text{S}^{-1}$)[64] and WS$_2$ ($130 \text{ cm}^2\text{V}^{-1}\text{S}^{-1}$)[65], demonstrating their fast carrier migration ability.

### 3.2.2: Optical Absorption Spectra

To investigate the light-harvesting ability of β-PdX$_2$ (X=S, Te) monolayer, we calculate the optical absorption spectra along the in-plane (XY) direction within the state-of-the-art GW+BSE method[66], which includes the exciton effects. β-PdX$_2$ (X=S, Te) monolayer structures exhibit strong optical absorption ability, as shown in Fig 3. Note that the optical absorbance (α(ω)) is plotted from the imaginary part of the dielectric function $\epsilon_2(\omega)$ on the basis of the following equation:[67, 68]

$$\alpha(\omega) = \frac{\omega}{c}\epsilon_2(\omega)L_z \tag{1}$$

where, $L_z$ represents the length of the supercell along the z-direction. The β-PdTe₂ monolayer exhibits relatively significant light absorption with prominent peaks in the visible region compared to β-PdS₂, as seen in Fig 3.

Also, the calculated excitons binding energy for β-PdS₂ and β-PdTe₂ monolayer is 0.84 eV and 0.75 eV, respectively, which is comparable to monolayer Cu₂ZnSnS₄ (0.84 eV)[69], Janus WSSe (0.83 eV)[70] and smaller than BeN₂ monolayer (1.07 eV)[71]. Hence β-PdX₂ (X=S, Te) monolayer has enough carriers to involve in the reaction with effectively separated electron-hole pairs.

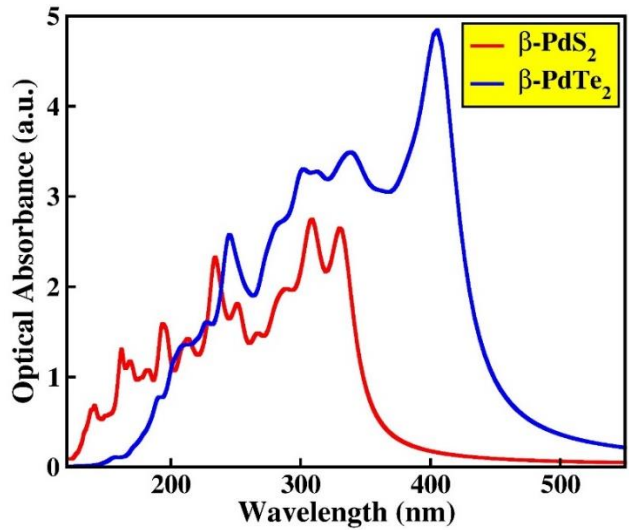

**Fig. 3** The absorbance of the β-PdX₂ (X= S, Te) monolayers computed using the GW + BSE method for the light incident in-plane (XY) direction.

### 3.3 Photocatalytic Properties

To assess whether the potential of utilizing the β-PdX₂ (X=S, Te) monolayer for photocatalytic water splitting, we further align the band edge positions with respect to vacuum level relative to that redox potential for hydrogen evolution ($E_{H^+/H_2}$) and oxygen evolution ($E_{O_2/H_2O}$) at pH = 0. The potential of the CBM (VBM) should be higher (lower) than the reduction level of hydrogen (oxidation level of oxygen) for an overall water splitting photocatalyst. As shown in Fig. 4(a), the VBM and CBM of β-PdS₂ monolayer surpass the standard oxidation ($O_2/H_2O$) and reduction ($H^+/H_2$) potential level, indicating sufficient activity of β-PdS₂ monolayer for both OER and HER processes. On the other hand, VBM of the β-PdTe₂ monolayer is larger than the oxidation level of oxygen (-5.67 at pH = 0) (Fig. 4(b)), indicating that oxidation half-reaction is not possible in β-PdTe$_2$ monolayer.

The external potentials produced by photogenerated carriers directly impact photocatalytic water splitting.[12, 72] The calculated potential of the photogenerated electrons for hydrogen reduction ($U_e$= energy difference between reduction potential of H$^+$/H$_2$ and the CBM) is 0.46 V at pH = 0 ($U_e$ = 0.46 +0.059 × pH) for β-PdS$_2$ monolayer. Whereas the potential of photogenerated holes of water oxidation ($U_h$= energy difference between reduction potential of H$^+$/H$_2$ and the VBM) is calculated to be 1.65 V at pH = 0 ($U_h$ = 1.65 – 0.059 × pH). Therefore, at a neutral environment (pH = 7), the $U_e$ and $U_h$ for the β-PdS$_2$ monolayer are 0.87 and 1.23 V, respectively, which indicates the potential to have photocatalytic activity water splitting in the neutral environment.

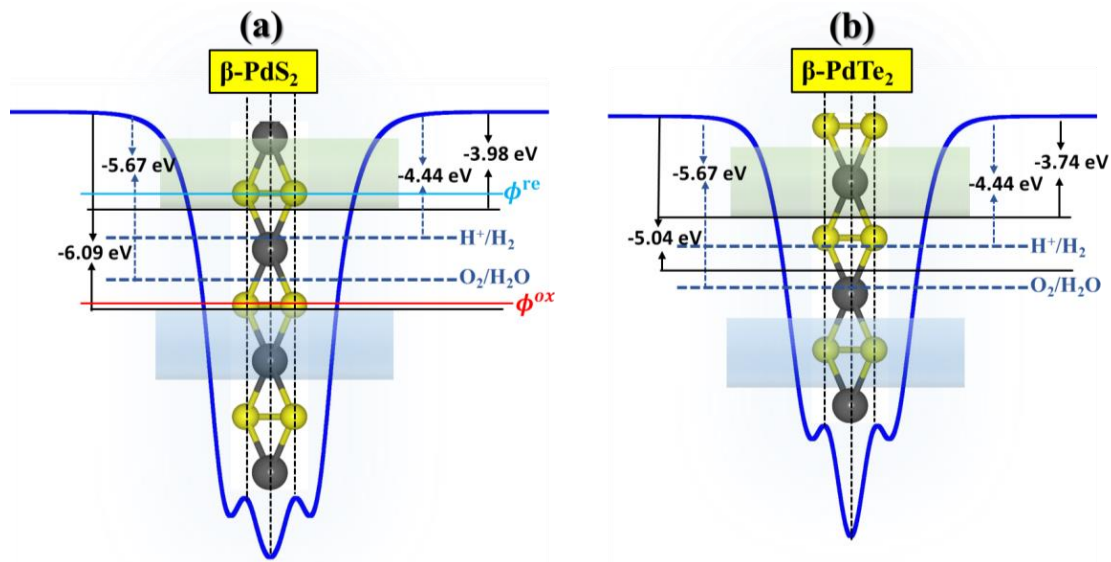

**Fig. 4** Band edge positions with water redox potential for water splitting at pH = 0 for (a) β-PdS$_2$ and (b) β-PdTe$_2$ monolayers along the z-direction based on the HSE06 level of theory.

To further assess the practical applications, the stability of photocatalysts in aqueous solution with illumination is also checked. Following Chen and Wang's method[73], we evaluate the thermodynamic oxidation ($\phi^{ox}$) (red line) and reduction potentials ($\phi^{re}$) (blue line) for β-PdS$_2$ monolayer (see the ESI). As displayed in Fig. 4a, the $\phi^{ox}$ are lower than the oxidation potential of O$_2$/H$_2$O; meanwhile, the $\phi^{re}$ are higher than the reduction potential of H$^+$/H$_2$, which reveals water molecules will be reduced and oxidized by photogenerated carriers instead of photocatalysts itself.[73] Hence β-PdS$_2$ monolayer have good resistance to photoinduced corrosion.

## 3.4 β-PdS$_2$ as Photocatalyst for Water Splitting

The presence of a suitable bandgap, band edge position and optical absorbance are not sufficient to imply the β-PdS$_2$ monolayer will act as a promising photocatalyst for overall water splitting. Therefore, to investigate further the water splitting activity, we examined the feasibility of adsorption and intercalation of water molecules on the surface of the β-PdS$_2$ monolayer, mechanisms and processes (Gibbs free energy) for both HER and OER and the solar to hydrogen (STH) efficiency for judging a material's ability to split water by photocatalysis.

### 3.4.1 Surface and Adsorption/Intercalation Energies

The adsorption and intercalation energies for the water molecules are calculated as $E_{ads/int} = E_{(*H2O)} − E_{(*)} − E_{(H2O)}$, where $E_{(*H2O)}$ represents the energy of (001) β-PdS$_2$ with adsorbed/ intercalated H$_2$O molecules, and $E_{(*)}$ is the energy of the pristine (001) β-PdS$_2$. Before estimating the adsorption/intercalation energy, we first check the stability of the (001) surface of β-PdS$_2$ by calculating the surface energy in the solvent. The computed surface energy (-0.13 eV) revealed that the (001) surface of β-PdS$_2$ is stable no matter in a vacuum or solution. The adsorbed/intercalated model is designed by a three-layer of β-PdS$_2$ (001) surface with an area of 60.2 Å$^2$ and adsorbed (Fig. 5a) or intercalated (Fig. 5b) with 1, 2, 3, 4 H$_2$O molecules. The adsorption and intercalation energy are calculated to be nearly the same (−0.65 eV/molecule and -0.66 eV/molecule) for various $H_2O$ coverage's (Table S4) reveal that β-$PdS_2$ (001) surface can equally adsorb or intercalate $H_2O$ molecules. We also estimate the energy for adsorption ($4H_2O$) with intercalation ($4H_2O$) model, as shown in Fig 5c, which result in lower energy (-0.35 eV per $H_2O$ molecules) than pure adsorption and intercalation energy. These results reveal that the coverage and incorporation of water molecules are energetically feasible, and water molecules can enter spontaneously to the β-$PdS_2$ layer.

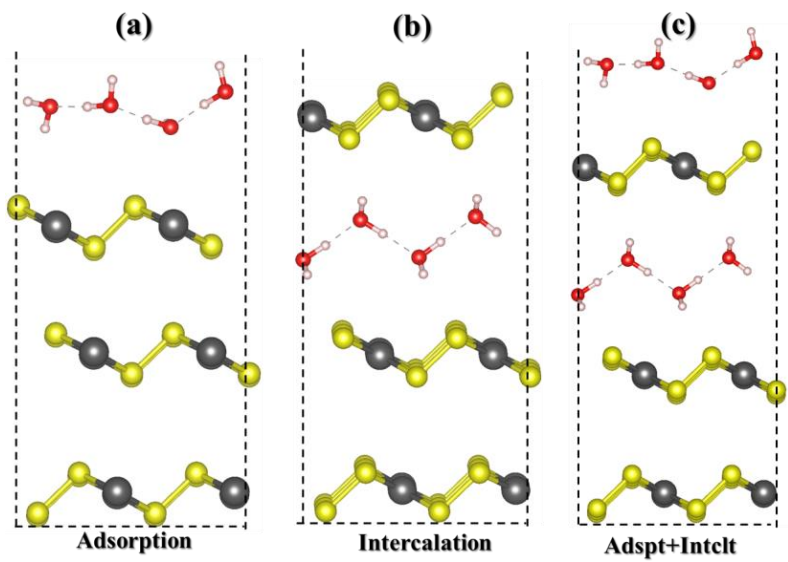

**Fig. 5** Adsorption and intercalation model: (a) 4 H$_2$O molecules on β-PdS$_2$ (001) surface; (b) 4 H$_2$O molecules intercalated in β-PdS$_2$ layers; (c) 4 H$_2$O molecules intercalated in layers and 4 H$_2$O molecules on (001) β-PdS$_2$ surface.

### 3.4.2 Gibbs Free Energy Profiles

Next, we consider the photocatalytic mechanism and Gibbs free energy profile of β-$PdS_2$ monolayer for HER and OER processes. The details of computations can be found in ESI. Note that we also include the solvation effect to describe the thermodynamics of chemical reactions at the solid/liquid interfaces. As displayed in Fig 6a, the hydrogen reduction only contains two steps.

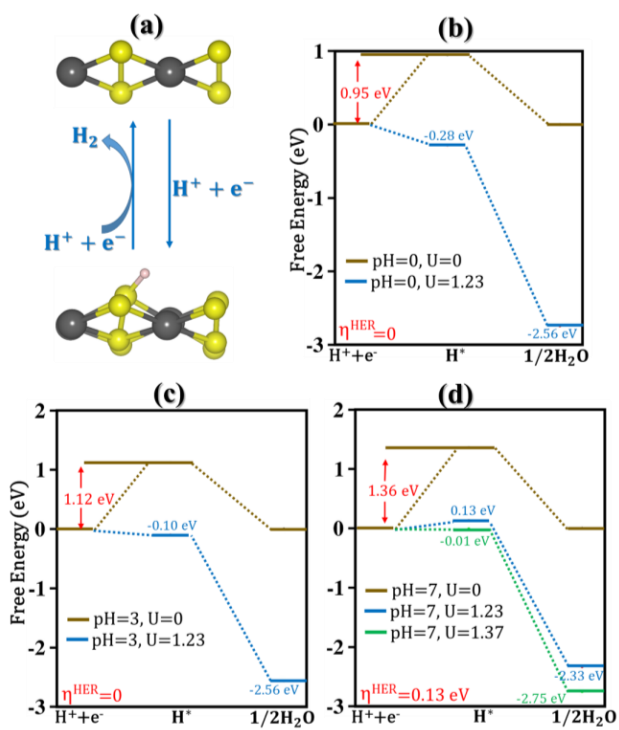

**Fig. 6** (a) Proposed photocatalytic pathways along with the atomic configuration of water oxidation. The white ball represents H atoms. The free-energy changes for HER at potential U = 0 and U = 1.23 V at (b) pH=0, (c) pH=3, and (d) pH=7.

First, the β-$PdS_2$ monolayer combines with a proton and an electron to form an H* species with an unfavorable ΔG of 0.95 eV (HER barrier) at U=0. Next, the $H_2$ molecule is released after the H* bonds a proton and an electron, exothermic by nature. However, when equilibrium potential of 1.23 V is considered at pH = 0-4, both the steps become downhill (Fig 6 and Fig S5 ESI), which indicates that the HER could happen spontaneously on the surface of the β-PdS$_2$ monolayer in an acid medium under illumination. At pH=7, an additional external potential of 0.13 V need to be employed to drive the HER to occur spontaneously (Fig 6d), which is similar to β-GeSe (0.13 V) [74] and much lower than β-AuS (0.16 V) [75], β-SnSe (0.16 V) [74], RhTeCl (0.30 V) [76], β-PdSe$_2$ (0.31 V) [77], C$_3$S (0.33 V) [78], SiP$_2$ (0.83 V), PE-AgBiP$_2$Se$_6$ (1.06 V) [72] and FE-AgBiP$_2$Se$_6$ (1.62 V) [72].

As illustrated in Fig 7a, the mechanism of water oxidation half-reaction following a four-electron reaction pathway accompanied by intermediate products. We compute the Gibbs free energy of intermediates (OH*, O*, OOH*) and ΔG of elementary steps to directly characterize the performance of the β-PdS$_2$ monolayer (Table S5). Initially, the adsorbed water molecule is

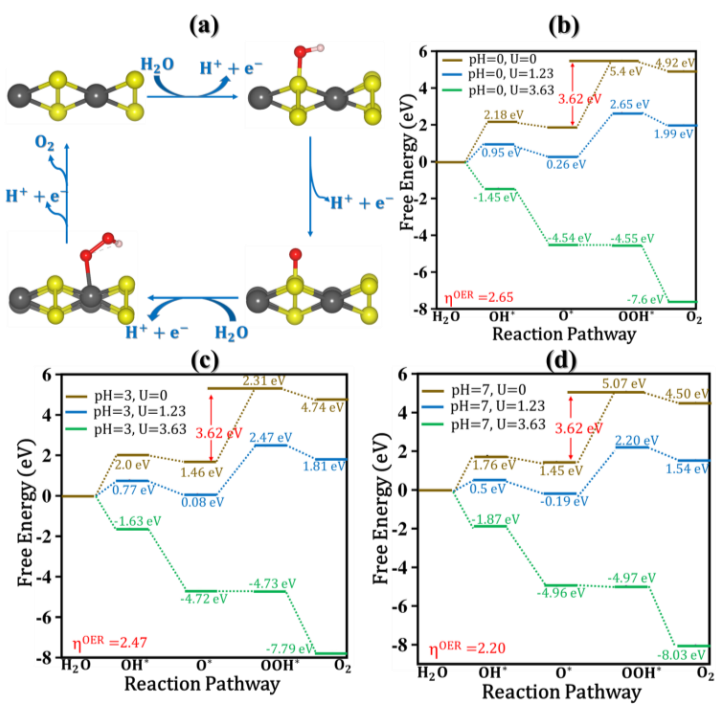

**Fig. 7** (a) Proposed photocatalytic pathways and the atomic configuration of absorbed intermediates species (OH*, O*, and OOH*) for water reduction. The free-energy changes of oxygen evolution at potential U = 0 and U = 1.23 V at (b) pH=0, (c) pH=3, and (d) pH=7. Note: the white and red balls represent H and O atoms, respectively.

oxidized into *OH species; second, the OH* species continues to be oxidized into O* intermediate after releasing another electron and a proton; third, combining with another water molecule, the *O species turns into an OOH* species; finally, the OOH* species is oxidized into free $O_2$ molecule by releasing one electron-proton pair. The third reaction step, OOH* formation, is the highest free energy change (OER barrier) with a limiting potential of 3.62 V (Fig 7b) in the water oxidation half-reaction. In the absence of any light irradiation (U = 0) at pH = 0−7 (brown line), the free energy change for step 3 ($\Delta G_{OOH*}$) is always uphill (endothermic), which means that the water oxidation half-reaction cannot proceed spontaneously (Fig 7 and Fig S6, ESI). Whereas, at the equilibrium potential of U=1.23 V (blue line), two reaction steps (first and third) were still endergonic at pH = 0-7.

At U=3.63 V (green line), the free energy profile revealed that all reaction steps were downhill. Thus, the computed OER overpotential ($\eta^{OER}$) became 2.65 V (2.47 V) for β-PdS$_2$ monolayer at pH=0 (pH=3), as shown in Fig. 7(b, c). Remarkably, the value of overpotential ($\eta^{OER}$) is reduced to 2.20 V (pH=7) from 2.65 V (pH=0), as shown in Fig. 7(d). Consequently, applying an external potential of 2.20 V can trigger the OER in a neutral medium.

### 3.4.3 Solar-to-Hydrogen (STH) Efficiency

The ultimate goal in the research of solar energy utilization for photocatalytic water splitting is to enhance energy conversion efficiency, which the solar-to-hydrogen efficiency could assess. Based on the band alignments, the energy conversion efficiencies of the β-PdS$_2$ monolayer are estimated with 100% efficiency of the catalytic reaction.[14] The calculated light absorption ($\eta_{abs}$), carrier utilization ($\eta_{cu}$), and STH ($\eta_{STH}$) efficiencies are listed in Table S6. The STH efficiency of photocatalytic water splitting for 2D material could be calculated as:

$$\eta_{STH} = \eta_{abs} \times \eta_{cu} \tag{5}$$

$$\eta_{STH} = \frac{\int_{E_g}^{\infty} P(\hbar\omega) d(\hbar\omega)}{\int_0^{\infty} P(\hbar\omega) d(\hbar\omega)} \times \frac{\Delta G \int_E^{\infty} \frac{P(\hbar\omega)}{\hbar\omega} d(\hbar\omega)}{\int_{E_g}^{\infty} P(\hbar\omega) d(\hbar\omega)} \tag{6}$$

Where $P(\hbar\omega)$ and $E$ represents the AM1.5G solar energy flux at the photon energy $\hbar\omega$ and energy of photons, respectively. (See ESI).

Here, the band gaps ($E_g$) and overpotentials of the hydrogen ($\chi(H_2)$), and oxygen ($\chi(O_2)$) are calculated using HSE06 hybrid functional (Table S6) to estimate these efficiencies. The light absorption efficiency ($\eta_{abs}$) of the β-PdS$_2$ monolayer is 32.35%, and it is highly dependent on the band gap value. Moreover, the energy conversion efficiencies of carrier utilization ($\eta_{cu}$) are higher than 35 % due to appropriate levels of $\chi(H_2)$ and $\chi(O_2)$ for a broad range of pH (0-7). The high efficiencies of both $\eta_{abs}$ and $\eta_{cu}$ lead to high STH efficiency. According to the definition of STH, $\eta_{STH}$ of β-PdS$_2$ monolayer are larger than 10% for all pH from 0 to 7 with 15.39 % highest for pH 3 and 4 (Fig. 8), which is much higher than that of pentagonal PdSe$_2$ (12.59%)[79], Ga$_2$S$_3$ (6.4%)[14], Ga$_2$SSe bilayer (7.42%)[80] and Al$_2$Se$_3$ (8%). Therefore, we believe that the β-PdS$_2$ monolayer meets the critical value for the economically production of hydrogen from photocatalysis (above 10%)[8].

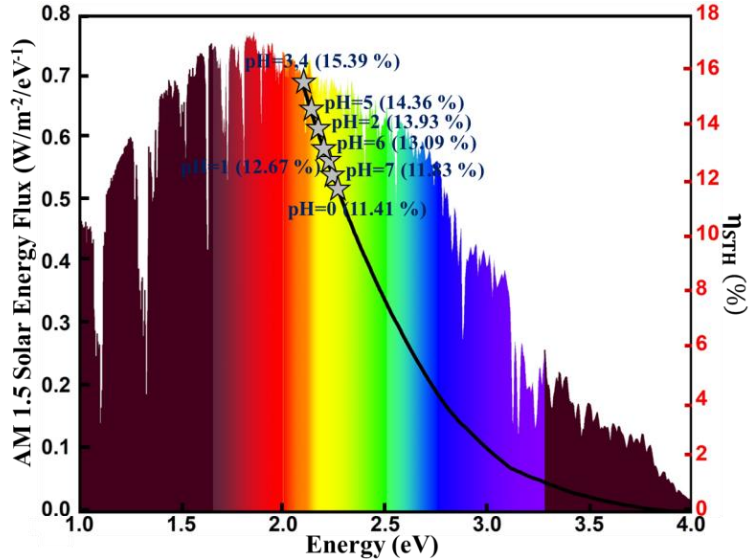

**Fig. 8** The solar energy photon flux of AM1.5G and theoretically predicted solar-to-hydrogen efficiency ($\eta_{STH}$) of the β-PdS$_2$ monolayer at different pH values at 100% quantum efficiency.

### 3.5 β-PdTe$_2$ as Excitonic Solar Cell

The suitable band gap, high carrier mobility and good solar light harvesting capability of β-PdTe$_2$ monolayer may be utilized to construct next-generation heterojunction solar cells. We consider eight appropriate and lattice-matching 2D materials (MoTe$_2$, WTe$_2$, Ga$_2$STe, InSe, InTe, RhTeCl, T-Te and β-PdS$_2$ monolayers) to build heterostructures with the β-PdTe$_2$ monolayer (Fig. S7 ESI). The electronic band structure of individual monolayers at the level of HSE06 level of theory is illustrated in Fig. S8 ESI. Moreover, the structural and electronic parameters such as the lattice constant, band gap, CBM, VBM, lattice mismatch, and interlayer distance with binding energies of these optimized configurations are summarized in Table S7. Figure 9a shows the CBM and VBM of these eight monolayers (acceptor) and β-PdTe$_2$ (donor). Encouragingly, the conduction band offset of β-PdTe$_2$, MoTe$_2$ and WTe$_2$ are calculated as 0.23 eV, 0.11 eV and 1 meV, resulting in high power conversion efficiency (PCE).

The PCE of β-PdTe$_2$/TMDs heterojunction solar cells is estimated by following the method of Scharber et al.[81] The upper limit of PCE at the 100% external quantum efficiency can be expressed as:

$$\eta = \frac{\beta_{FF}V_{oc}J_{sc}}{P_{solar}} = \frac{0.65(E_g^d - \Delta E_c - 0.3)\int_{E_g^d}^{\infty}\frac{P(\hbar\omega)}{\hbar\omega}d(\hbar\omega)}{\int_0^{\infty}P(\hbar\omega)d(\hbar\omega)}$$

Here $P(\hbar\omega)$ is the AM1.5 solar energy flux (expressed in Wm$^2$eV$^{-1}$) at the photon energy $\hbar\omega$, $E_g$ is the band gap of the donor obtained from HSE06 calculations, and 0.65 is the band-bill factor deduced from Shockley– Queisser limit.[82] The $J_{sc}$ is the short circuit current assuming an external quantum efficiency of 100% and $\Delta E_c$ is CBO between the donor and acceptor.

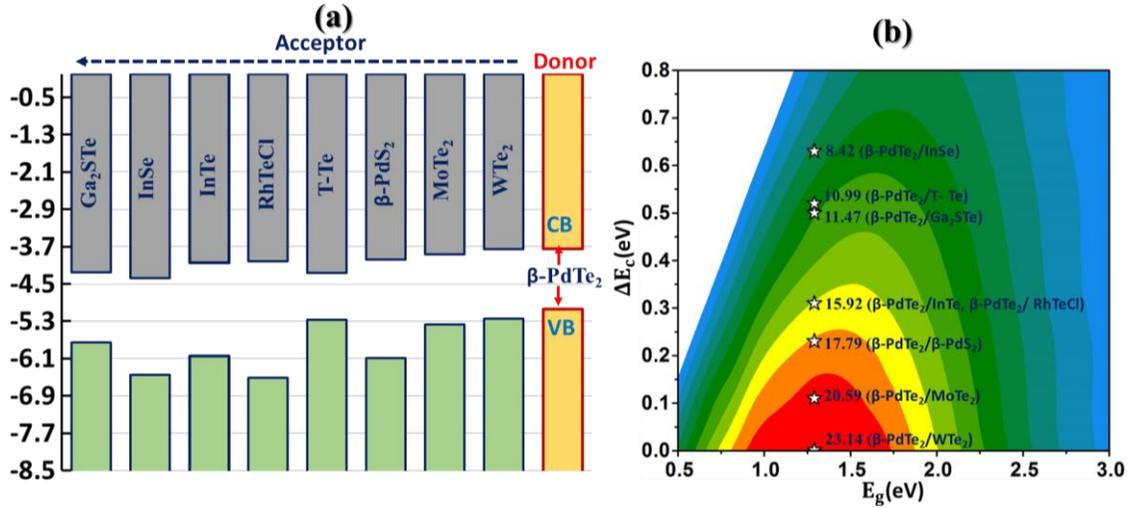

**Fig. 9** (a) Band alignments of β-PdTe$_2$ with other 2D materials (MoTe$_2$, WTe$_2$, Ga$_2$STe, InSe, InTe, RhTeCl, T-Te and β-PdS$_2$) monolayers obtained by HSE06 functional level of theory with vacuum level correction. (b) Computed exciton solar energy efficiency contour as a function of the CBO and donor band gap (E$_g$) of the designed vdW heterostructures.

The PCEs of β-PdTe$_2$/TMDs heterojunction solar cells are shown in Fig. 9b (Table S8). As donor materials, β-PdTe$_2$ has a band gap of 1.29 eV, which is suitable for absorbing the solar spectrum. Also, the combined effects of the matched donor-acceptor band alignments lead to lower conduction band offset, resulting in high efficiency. The PCEs of β-PdTe$_2$/β-PdS$_2$, β-PdTe$_2$/ MoTe$_2$ and β-PdTe$_2$/WTe$_2$ heterojunctions are calculated as 17.79%, 20.59% and 23.14%, respectively.

Table 1 lists the PCEs of vdW heterostructures and other 2D heterostructure solar cells for a more intuitive comparison. Our proposed β-PdTe$_2$/β-PdS$_2$, β-PdTe$_2$/ MoTe$_2$ and β-PdTe$_2$/WTe$_2$ heterojunctions have a competitive edge over existing 2D heterojunction solar cells.

**Table 1** Power conversion efficiency (PCE) of recently proposed 2D heterojunction solar cells.

| Heterojunctions | PCE (%) | Ref. |
|---|---|---|
| β-PdTe$_2$/ WTe$_2$, β-PdTe$_2$/ MoTe$_2$, β-PdTe$_2$/ β-PdS$_2$ | 23.14, 20.59, 17.79 | This work |
| Te/WTe$_2$, Te/ MoTe$_2$ | 22.5, 20.1 | 83 |
| Ti$_2$CO$_2$/Zr$_2$CO$_2$, Ti$_2$CO$_2$/Hf$_2$CO$_2$ | 22.74, 19.56 | 84 |
| HfTeSe$_4$/Bi$_2$WO$_6$ | 20.8 | 76 |
| TiNF/TiNCl, TiNCl/TiNBr | 22, 19 | 29 |
| MoS2/ψ-phosphorene | 20.26 | 85 |
| α-AsP/GaN | 22.1 | 86 |
| BP/MoSSe, BaS/MoSSe | 22.97, 20.86 | 87 |
| Sc$_2$COHH/InS | 21.04 | 88 |
| P-PdSe$_2$/MoSe$_2$, P-PdSe$_2$/MoTe$_2$ | 22, 17 | 89 |
| Pb$_2$SSe/SnSe, Pb$_2$SSe/GeSe | 20.02, 19.28 | 90 |

**4. Conclusions**

In summary, using first-principles calculations, we propose the 2D β-PdX$_2$ (X=S, Te) monolayers for efficient solar energy conversion applications. We systematically explored the stability, optoelectronic, photocatalytic and photovoltaic properties of the newly emerging β-PdX$_2$ (X=S, Te) monolayers. The thermal and kinetic stability evaluated based on AIMD simulations and phonon dispersion, and the energetics based on cleavage and cohesive energy suggests their feasibility for experimental synthesis. The moderate carrier mobility and pronounced light absorption ability of the β-PdX$_2$ monolayer make them a suitable candidate for energy conversion devices. The β-PdS$_2$ monolayer shows a considerable band gap of 2.10 eV, and redox potential is within band edge position of VBM and CBM. Especially, the photogenerated holes have adequate driving forces to render the hydrogen reduction half-reactions proceed spontaneously on the β-PdS$_2$ monolayer at pH= 0 to 4. Besides, the additional external potential of 2.20 V for OER and 0.13 V for HER would be needed to trigger the whole sequence for β-PdS$_2$ monolayer at pH=7.

Moreover, the adsorption/intercalation energies of β-PdS$_2$ revealed the coverage and incorporation of water molecules are energetically feasible with high STH efficiency. In addition, the 2D β-PdTe$_2$, as characterized by a moderate bandgap of 1.29 eV, and suitable donor material for constructing heterostructure as exciton solar cell. The proposed β-PdTe$_2$/β-PdS$_2$, β-PdTe$_2$/ MoTe$_2$ and β-PdTe$_2$/WTe$_2$ heterojunctions can achieve a power conversion efficiency (PCE) up to 17.79%, 20.59% and 23.14%, respectively. It is expected that these new 2D β-PdS$_2$ and β-PdS$_2$ monolayers can be applied in photocatalysis and photovoltaic in the near future.

## Supporting Information

Detailed information of lattice parameters for monolayer and bulk, electronic, carrier mobility, STH efficiency, Gibbs free energy mechanism and heterostructure for β-PdX$_2$ (X= S, Te) monolayers.

## Acknowledgments

The computational facility available at the Central University of Punjab, Bathinda, was used to obtain the results presented in this paper. MJ is thankful to the University Grants Commission (UGC) for financial assistance in the form of the Senior Research Fellowship. Helpful discussion with Jaspreet Singh is highly acknowledged.

## References

1. A. Fujishima and K. Honda, *nature*, 1972, **238**, 37-38.
2. J. Suntivich, K. J. May, H. A. Gasteiger, J. B. Goodenough and Y. Shao-Horn, *Science*, 2011, **334**, 1383-1385.
3. R. Asahi, T. Morikawa, T. Ohwaki, K. Aoki and Y. Taga, *science*, 2001, **293**, 269-271.
4. T. Yanagida, Y. Sakata and H. Imamura, *Chemistry letters*, 2004, **33**, 726-727.
5. K. Maeda, T. Takata, M. Hara, N. Saito, Y. Inoue, H. Kobayashi and K. Domen, *Journal of the American Chemical Society*, 2005, **127**, 8286-8287.
6. N. Han, P. Liu, J. Jiang, L. Ai, Z. Shao and S. Liu, *Journal of Materials Chemistry A*, 2018, **6**, 19912-19933.
7. S. Sun, T. Hisatomi, Q. Wang, S. Chen, G. Ma, J. Liu, S. Nandy, T. Minegishi, M. Katayama and K. Domen, *ACS Catalysis*, 2018, **8**, 1690-1696.
8. C. R. Cox, J. Z. Lee, D. G. Nocera and T. Buonassisi, *Proceedings of the National Academy of Sciences*, 2014, **111**, 14057-14061.
9. G. Zhang, Z.-A. Lan, L. Lin, S. Lin and X. Wang, *Chemical science*, 2016, **7**, 3062-3066.

10. A. Fujishima, T. N. Rao and D. A. Tryk, *Journal of photochemistry and photobiology C: Photochemistry reviews*, 2000, **1**, 1-21.
11. H. Taghinejad, D. A. Rehn, C. Muccianti, A. A. Eftekhar, M. Tian, T. Fan, X. Zhang, Y. Meng, Y. Chen and T.-V. Nguyen, *Acs Nano*, 2018, **12**, 12795-12804.
12. M. Qiao, J. Liu, Y. Wang, Y. Li and Z. Chen, *Journal of the American Chemical Society*, 2018, **140**, 12256-12262.
13. Y. Wan, L. Wang, H. Xu, X. Wu and J. Yang, *Journal of the American Chemical Society*, 2020, **142**, 4508-4516.
14. C.-F. Fu, J. Sun, Q. Luo, X. Li, W. Hu and J. Yang, *Nano letters*, 2018, **18**, 6312-6317.
15. X. Zhang, Z. Zhang, D. Wu, X. Zhang, X. Zhao and Z. Zhou, *Small Methods*, 2018, **2**, 1700359.
16. A. K. Singh, K. Mathew, H. L. Zhuang and R. G. Hennig, *The journal of physical chemistry letters*, 2015, **6**, 1087-1098.
17. Y. Li, Y.-L. Li, B. Sa and R. Ahuja, *Catalysis Science & Technology*, 2017, **7**, 545-559.
18. L. Wang, Y. Wan, Y. Ding, S. Wu, Y. Zhang, X. Zhang, G. Zhang, Y. Xiong, X. Wu and J. Yang, *Advanced Materials*, 2017, **29**, 1702428.
19. P. Jamdagni, R. Pandey and K. Tankeshwar, *Nanotechnology*, 2021, **33**, 025703.
20. G. Gao, Y. Jiao, F. Ma, Y. Jiao, E. Waclawik and A. Du, *The Journal of Physical Chemistry C*, 2015, **119**, 13124-13128.
21. H. Huang, X. Fan, D. J. Singh and W. Zheng, *Nanoscale*, 2020, **12**, 1247-1268.
22. S. Manzeli, D. Ovchinnikov, D. Pasquier, O. V. Yazyev and A. Kis, *Nature Reviews Materials*, 2017, **2**, 1-15.
23. B. Luo, G. Liu and L. Wang, *Nanoscale*, 2016, **8**, 6904-6920.
24. A. S. Bati, M. Batmunkh and J. G. Shapter, *Advanced Energy Materials*, 2020, **10**, 1902253.
25. L. Wang, L. Huang, W. C. Tan, X. Feng, L. Chen, X. Huang and K. W. Ang, *Small Methods*, 2018, **2**, 1700294.
26. T. Su, Q. Shao, Z. Qin, Z. Guo and Z. Wu, *Acs Catalysis*, 2018, **8**, 2253-2276.
27. D. M. Chapin, C. S. Fuller and G. L. Pearson, *Journal of Applied Physics*, 1954, **25**, 676-677.
28. J. Linghu, T. Yang, Y. Luo, M. Yang, J. Zhou, L. Shen and Y. P. Feng, *ACS applied materials & interfaces*, 2018, **10**, 32142-32150.
29. Y. Liang, Y. Dai, Y. Ma, L. Ju, W. Wei and B. Huang, *Journal of Materials Chemistry A*, 2018, **6**, 2073-2080.
30. K. S. Novoselov, A. K. Geim, S. V. Morozov, D.-e. Jiang, Y. Zhang, S. V. Dubonos, I. V. Grigorieva and A. A. Firsov, *science*, 2004, **306**, 666-669.
31. K. S. Novoselov, A. K. Geim, S. V. Morozov, D. Jiang, M. I. Katsnelson, I. Grigorieva, S. Dubonos, Firsov and AA, *nature*, 2005, **438**, 197-200.
32. M. J. Allen, V. C. Tung and R. B. Kaner, *Chemical reviews*, 2010, **110**, 132-145.

33. C. Xia, J. Du, W. Xiong, Y. Jia, Z. Wei and J. Li, *Journal of Materials Chemistry A*, 2017, **5**, 13400-13410.

34. Y.-L. Liu, Y. Shi and C.-L. Yang, *Applied Surface Science*, 2021, **545**, 148952.

35. P. Giannozzi, O. Andreussi, T. Brumme, O. Bunau, M. B. Nardelli, M. Calandra, R. Car, C. Cavazzoni, D. Ceresoli and M. Cococcioni, *Journal of physics: Condensed matter*, 2017, **29**, 465901.

36. P. Giannozzi, S. Baroni, N. Bonini, M. Calandra, R. Car, C. Cavazzoni, D. Ceresoli, G. L. Chiarotti, M. Cococcioni and I. Dabo, *Journal of physics: Condensed matter*, 2009, **21**, 395502.

37. J. Heyd, G. E. Scuseria and M. Ernzerhof, *The Journal of chemical physics*, 2003, **118**, 8207-8215.

38. A. V. Krukau, O. A. Vydrov, A. F. Izmaylov and G. E. Scuseria, *The Journal of chemical physics*, 2006, **125**, 224106.

39. S. Grimme, *Journal of computational chemistry*, 2006, **27**, 1787-1799.

40. J. M. Soler, E. Artacho, J. D. Gale, A. García, J. Junquera, P. Ordejón and D. Sánchez-Portal, *Journal of Physics: Condensed Matter*, 2002, **14**, 2745.

41. A. Marini, C. Hogan, M. Grüning and D. Varsano, *Computer Physics Communications*, 2009, **180**, 1392-1403.

42. O. Andreussi, I. Dabo and N. Marzari, *The Journal of chemical physics*, 2012, **136**, 064102.

43. D. K. Sang, T. Ding, M. N. Wu, Y. Li, J. Li, F. Liu, Z. Guo, H. Zhang and H. Xie, *Nanoscale*, 2019, **11**, 18116-18123.

44. N. Mounet, M. Gibertini, P. Schwaller, D. Campi, A. Merkys, A. Marrazzo, T. Sohier, I. E. Castelli, A. Cepellotti and G. Pizzi, *Nature nanotechnology*, 2018, **13**, 246-252.

45. K. F. Mak, C. Lee, J. Hone, J. Shan and T. F. Heinz, *Physical review letters*, 2010, **105**, 136805.

46. Y. Guo, Y. Zhang, S. Yuan, B. Wang and J. Wang, *Nanoscale*, 2018, **10**, 18036-18042.

47. R. Zacharia, H. Ulbricht and T. Hertel, *Physical Review B*, 2004, **69**, 155406.

48. S. Zhao, Z. Li and J. Yang, *Journal of the American Chemical Society*, 2014, **136**, 13313-13318.

49. D. L. Druffel, K. L. Kuntz, A. H. Woomer, F. M. Alcorn, J. Hu, C. L. Donley and S. C. Warren, *Journal of the American Chemical Society*, 2016, **138**, 16089-16094.

50. C. Liu, G. Zhao, T. Hu, L. Bellaiche and W. Ren, *Physical Review B*, 2021, **103**, L081403.

51. S. Sun, F. Meng, H. Wang, H. Wang and Y. Ni, *Journal of Materials Chemistry A*, 2018, **6**, 11890-11897.

52. M. Qiao, Y. Chen, Y. Wang and Y. Li, *Journal of Materials Chemistry A*, 2018, **6**, 4119-4125.

53. X. Zhang, J. Liu, E. Zhang, R. Pan, Y. Li, X. Wan, H. Wang and J. Zhang, *Chemical Communications*, 2020, **56**, 5504-5507.

54. M. Campetella and M. Calandra, *Physical Review B*, 2021, **104**, 174111.

55. R. Tian, A. Griffin, M. McCrystall, M. Breshears, A. Harvey, C. Gabbett, D. V. Horváth, C. Backes, Y. Jing and T. Heine, *Advanced Energy Materials*, 2021, **11**, 2002364.

56. P. Zhang, F. Zhao, P. Long, Y. Wang, Y. Yue, X. Liu, Y. Feng, R. Li, W. Hu and Y. Li, *Nanoscale*, 2018, **10**, 15989-15997.
57. T. Björkman, A. Gulans, A. V. Krasheninnikov and R. M. Nieminen, *Physical review letters*, 2012, **108**, 235502.
58. D. Fan, S. Lu, C. Chen, M. Jiang, X. Li and X. Hu, *Applied Physics Letters*, 2020, **117**, 013103.
59. Y. Jing, Y. Ma, Y. Li and T. Heine, *Nano letters*, 2017, **17**, 1833-1838.
60. N. Miao, B. Xu, N. C. Bristowe, J. Zhou and Z. Sun, *Journal of the American Chemical Society*, 2017, **139**, 11125-11131.
61. L.-M. Yang, I. A. Popov, T. Frauenheim, A. I. Boldyrev, T. Heine, V. Bačić and E. Ganz, *Physical Chemistry Chemical Physics*, 2015, **17**, 26043-26048.
62. A. D. Oyedele, S. Yang, L. Liang, A. A. Puretzky, K. Wang, J. Zhang, P. Yu, P. R. Pudasaini, A. W. Ghosh and Z. Liu, *Journal of the American Chemical Society*, 2017, **139**, 14090-14097.
63. J. Bardeen and W. Shockley, *Physical review*, 1950, **80**, 72.
64. K. Kaasbjerg, K. S. Thygesen and K. W. Jacobsen, *Physical Review B*, 2012, **85**, 115317.
65. A. Rawat, N. Jena and A. De Sarkar, *Journal of Materials Chemistry A*, 2018, **6**, 8693-8704.
66. M. Rohlfing and S. G. Louie, *Physical Review B*, 2000, **62**, 4927.
67. A. Kumar, G. Sachdeva, R. Pandey and S. P. Karna, *Applied Physics Letters*, 2020, **116**, 263102.
68. J. Singh, P. Jamdagni, M. Jakhar and A. Kumar, *Physical Chemistry Chemical Physics*, 2020, **22**, 5749-5755.
69. R. Zhang, X. Wen, F. Xu, Q. Zhang and L. Sun, *The Journal of Physical Chemistry C*, 2020, **124**, 11922-11929.
70. L. Ju, M. Bie, X. Tang, J. Shang and L. Kou, *ACS Applied Materials & Interfaces*, 2020, **12**, 29335-29343.
71. Y.-m. Ding, Y. Ji, H. Dong, N. Rujisamphan and Y. Li, *Nanotechnology*, 2019, **30**, 465202.
72. L. Ju, J. Shang, X. Tang and L. Kou, *Journal of the American Chemical Society*, 2019, **142**, 1492-1500.
73. S. Chen and L.-W. Wang, *Chemistry of Materials*, 2012, **24**, 3659-3666.
74. Y. Xu, K. Xu, C. Ma, Y. Chen, H. Zhang, Y. Liu and Y. Ji, *Journal of Materials Chemistry A*, 2020, **8**, 19612-19622.
75. L. Lv, Y. Shen, X. Gao, J. Liu, S. Wu, Y. Ma, X. Wang, D. Gong and Z. Zhou, *Applied Surface Science*, 2021, **546**, 149066.
76. H. Yang, Y. Ma, Y. Liang, B. Huang and Y. Dai, *ACS applied materials & interfaces*, 2019, **11**, 37901-37907.
77. M. Jakhar and A. Kumar, *Catalysis Science & Technology*, 2021, **11**, 6445-6454.
78. M. Tang, B. Wang, H. Lou, F. Li, A. Bergara and G. Yang, *The Journal of Physical Chemistry Letters*, 2021, **12**, 8320-8327.

79. C. Long, Y. Liang, H. Jin, B. Huang and Y. Dai, *ACS Applied Energy Materials*, 2018, **2**, 513-520.
80. Y. Bai, R. Guan, H. Zhang, Q. Zhang and N. Xu, *Catalysis Science & Technology*, 2021, **11**, 542-555.
81. M. C. Scharber, D. Mühlbacher, M. Koppe, P. Denk, C. Waldauf, A. J. Heeger and C. J. Brabec, *Advanced materials*, 2006, **18**, 789-794.
82. Y. Xu, T. Gong and J. N. Munday, *Scientific reports*, 2015, **5**, 1-9.
83. K. Wu, H. Ma, Y. Gao, W. Hu and J. Yang, *Journal of Materials Chemistry A*, 2019, **7**, 7430-7436.
84. Y. Zhang, R. Xiong, B. Sa, J. Zhou and Z. Sun, *Sustainable Energy & Fuels*, 2021, **5**, 135-143.
85. H. Wang, X. Li, Z. Liu and J. Yang, *Physical Chemistry Chemical Physics*, 2017, **19**, 2402-2408.
86. M. Xie, S. Zhang, B. Cai, Y. Huang, Y. Zou, B. Guo, Y. Gu and H. Zeng, *Nano Energy*, 2016, **28**, 433-439.
87. M. K. Mohanta and A. De Sarkar, *Nanoscale*, 2020, **12**, 22645-22657.
88. Y. Zhang, B. Sa, N. Miao, J. Zhou and Z. Sun, *Journal of Materials Chemistry A*, 2021, **9**, 10882-10892.
89. M. Jakhar, J. Singh, A. Kumar and R. Pandey, *The Journal of Physical Chemistry C*, 2020, **124**, 26565-26571.
90. F. Zhang, J. Qiu, H. Guo, L. Wu, B. Zhu, K. Zheng, H. Li, Z. Wang, X. Chen and J. Yu, *Nanoscale*, 2021, **13**, 15611-15623.